\documentclass[aps,pre,twocolumn,floats,amssymb]{revtex4}

\usepackage{graphicx}
\usepackage{amsmath}
\usepackage{amssymb}
\bibstyle{apsrev.bib}

\usepackage{booktabs}

\begin{document}

\title{Totally Asymmetric Exclusion Process with Hierarchical Long-Range Connections}

\author{Jakub Otwinowski and Stefan Boettcher}
\email{www.physics.emory.edu/faculty/boettcher}
\affiliation{Physics Department, Emory University, Atlanta, GA 30322, USA}

\date{\today}
\begin{abstract}
A non-equilibrium particle transport model, the totally asymmetric
exclusion process, is studied on a one-dimensional lattice with a
hierarchy of fixed long-range connections. This model breaks the
particle-hole symmetry observed on an ordinary one-dimensional lattice
and results in a surprisingly simple phase diagram, without a
maximum-current phase. Numerical simulations of the model with open
boundary conditions reveal a number of dynamic features and suggest
possible applications.
\end{abstract}
\maketitle

\section{Introduction}
\label{introduction}
Physicists have long hoped to understand non-equilibrium phenomena as
well as they understand equilibrium phenomena \citep{Schmittmann95}.
Certain non-equilibrium systems exist which reach a steady state, yet
they do not obey detailed balance required for any equilibrium.  The
steady states are defined by the dynamics rather than an energy
function. Exclusion processes are widely studied as models of particle
transport and were first introduced for the kinetics of
bio-polymerization on nucleic acid templates
\citep{citeulike:1532138}. They have since been related to other
phenomena such as surface growth \citep{Barabasi95}, traffic flow
\citep{Biham92,Nagel92,Nagel95}, and the statistics of DNA sequence
alignment \citep{citeulike:3175269}. The totally asymmetric exclusion
process (TASEP) represents a rare example of an exactly solvable model
with a non-equilibrium steady state that allows a deep, analytic
insight into the phenomenology of critical behavior beyond
thermodynamic equilibrium
\citep{citeulike:1341557,citeulike:3196043,citeulike:3196051}. TASEP
and related models exhibit non-equilibrium phase transitions which
have no analog in equilibrium systems, such as a phase transition in
one dimension.

TASEP describes particles conducting nearest neighbor jumps along one
direction within a line of sites. There is no passing, and jumps of
particles to an occupied forward site are excluded, leading to
jamming. The model has been solved first in a mean-field treatment
\citep{citeulike:1341557} and subsequently in full detail
\citep{citeulike:3196043,citeulike:3196051}, and has since inspired a
large number of variations \citep{Hinrichsen00,Mobilia08}.  Yet, the
phase diagram has proven quite robust under those changes.  Even
admitting long-range jumps, which allows particles to pass forward by
a stochastic long-range jump according to a Levy distribution if the
target site is unoccupied, leaves the phase diagram surprisingly
unchanged \citep{citeulike:3163805}.

The purpose of this paper is to introduce jumps into the TASEP in a
non-stochastic way by using a network with predetermined long-distance
jumps and to study it's effect on the phases and transitions. Such a
quenched structure is provided by the recently introduced network
HN3. HN3 has a hierarchical structure, combining a simple
one-dimensional backbone with a sequence of long-range
links. Interesting properties for other statistical models on HN3 have
already been described in Ref.~\citep{SWPRL}. The process on HN3
(HN3-TASEP) might prove to be a benign enough extension beyond that in
one dimension ($1d$-TASEP) such that analytical insights remain
possible, even though introducing quenched long-range connections
removes the particle-hole symmetry. Here, we show numerically that HN3
alters the phase diagram significantly. In fact, only two phases
remain, separated by a sharp first-order transition. Our results
suggest that there might indeed be a simple solution for this model
which would take the analytic treatment of exclusion processes beyond
one dimension. The observed behavior suggests that the hierarchical
lattice used here also provides an efficient switch for a simple, one
parameter storage-and-release system.

There are some real systems to which an exclusion process with
particles passing each other applies. For instance, as
pointed out in Ref.~\citep{citeulike:3175269}, some proteins regulate
genes by binding to DNA and search for a specific target site by
continuous dissociation and re-association with the DNA. In the
dissociation process, they may fully dissociate or stay within the
range of electrostatic forces of the DNA. In another application,
HN3-TASEP could model traffic \citep{Biham92,Nagel92,Nagel95}
with expressways, which also function as quenched shortcuts. The
hierarchical but geometric structure of HN3 may even  apply to
a multi-level transport system such as a package delivery service with
many door-to-door vans, a number of intercity trucks and trains, and a
few transcontinental flight routes.

Our discussion proceeds with a review of the ordinary $1d$-TASEP in
the next section, followed in Sec.~\ref{sec:TaSEP-on-HN3} by a
description of the hierarchical lattice geometry employed here. In
Secs.~\ref{sec:Numerical-Results} and ~\ref{Sec:Analytic}, we present
our numerical results and analytic approach, respectively. In Sec.~\ref{sec:Discussion} we discuss the implications of our
findings and
provide our conclusions in Sec.~\ref{sec:Conclusions}.

\section{TASEP on a line\label{sec:TASEP-on-a line}}
\begin{figure}
\vskip 2.2in
\includegraphics{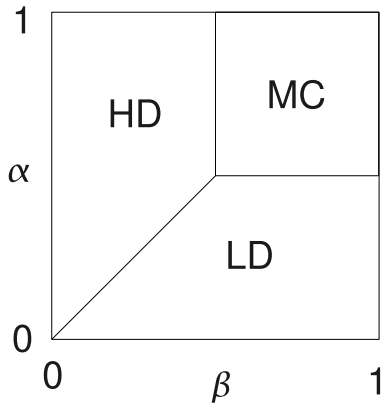}
\caption{\label{fig:Phasediagram-standard}
Phase diagram of the $1d$-TASEP with high density (HD), low density
(LD) and maximum current (MC) phases.  The injection rate is $\alpha$ 
and the removal rate is $\beta$.}
\end{figure}

The totally asymmetric exclusion process in one dimensional
($1d$-TASEP), where particles always move in one direction, is a model
which has several phases with first and second-order transitions
\citep{citeulike:1341557,citeulike:3196043,citeulike:3196051}.  It is
defined on a one-dimensional lattice of length $L$. Each site of the
lattice, labeled by $i$, is either occupied or unoccupied by a
particle, and accordingly has an occupation number $\tau_{i}$ which is
either 1 or 0. Particles on the lattice may only move in one
direction, which we will say is to the right, and they may hop one
site to the right only if that site is unoccupied. Typically, the
system is updated random sequentially, that is, particles are selected
to move one at a time in random order. The average density $\rho$ is
defined as $N/L$, where $N$ is the number of particles in the
system. The average current $J$ is defined as the average number of
particles that move through a point on the lattice per time step. For
periodic boundary conditions, $N$ is fixed, while for open boundary
conditions, as considered here, $N$ is allowed to fluctuate,
which induces several phases characterized by different properties for
$\rho$ and $J$. The two open boundaries are connected to a large
reservoir of particles, so that the rate of particles hopping onto the
lattice at the first site is $\alpha,$ and the rate of particles
removed from the last site is $\beta$, providing two independent
parameters. Both define a parameter space for which the different
phases appear, see Fig.~\ref{fig:Phasediagram-standard}.
Their rates are between 0 and 1, as no more than one particle may
appear or disappear on the boundary in a single trial.

Fig.~\ref{fig:Phasediagram-standard} shows three phases distinguished
by their average densities and currents in the steady state, first
found in Ref.~\citep{citeulike:1341557}. The low density (LD) phase
has $\rho_{LD}=\alpha$ and $J_{LD}=\alpha(1-\alpha)$, the high density
(HD) phase has $\rho_{HD}=1-\beta$ and $J_{HD}=\beta(1-\beta)$, and
the maximum current (MC) phase has $\rho_{MC}=1/2$ and $J_{MC}=1/4$.
The HD and LD phases are related because of particle-hole symmetry.
The transitions between the LD and HD phases and the MC phase are
continuous. The line between the LD and HD phases is called the
{}``shock phase'' (SP). This transition is not continuous and the
system reaches a state of coexistence between the two phases. On the
lattice there is a region with low density and a region of high
density separated by microscopically small {}``shock'', which diffuses
between the ends of the lattice.

$1d$-TASEP has been solved exactly with recursive equations
\citep{citeulike:1341557,citeulike:3196051}, as well as a more
advanced matrix formulation \citep{citeulike:3196043}.  The process is
completely described by the change in occupations on affected sites
for an update at a bulk-site $i$ during the time interval
$\left[t,t+dt\right]$. Such an update merely alters site $i$ and
$i+1$:
\begin{eqnarray}
\tau_{i}(t+dt)&=&\tau_{i}(t)\tau_{i+1}(t)\label{eq:master2}\\
\tau_{i+1}(t+dt)&=&\tau_{i+1}(t)+\left[1-\tau_{i+1}(t)\right]\tau_{i}(t),
\nonumber
\end{eqnarray}
all other sites remain unchanged.  Special treatment obtains for each
of the two boundary sites. When the update selects to inject a new
particle into the system, that particle attempts to occupy site $i=1$
with probability $\alpha$, if that site is open. When the last site,
$i=L$, is selected for an update, it unloads an occupying particle
with probability $\beta$.

These master equations, as well as those boundary conditions, can be
averaged over noise, eliminating all fluctuations, and written as
single differential equation describing the time evolution of the
system. Assuming the existence of a steady state,
$\partial_{t}\left\langle \tau_{i}\right\rangle =0$, leads to a system
of algebraic equations at most quadratic in $\left\langle
\tau_{i}\right\rangle $, which can be solved recursively.

Many variations of TASEP have been studied, such as those with
parallel updates, multiple species of particles, and extended particle
sizes \citep{Mobilia08}. Remarkably, the phase diagram is essentially
the same for many variations. For example, one variation studied
recently introduced long range jumps so that particles may jump past
each other (if the target site is unoccupied) to study the changes to
the phases and transitions \citep{citeulike:3163805}. These jumps are
stochastic Levy flights, i.~e. a distance $l$ is reached with
probability $p_{l}\sim l^{-(1+\sigma)}$ depending on a parameter
$\sigma>0$. Even with long jumps and reordering ({}``passing'') of
particles, the diagram has the same three phases as in
Fig.~\ref{fig:Phasediagram-standard}.

\section{TASEP on the Hierarchical Network HN3\label{sec:TaSEP-on-HN3}}
HN3 is a network with a fractional dimension, which was introduced
with the intention of studying small-world phenomena analytically
\citep{SWPRL}. Small-world phenomena are found in many natural and
man-made systems, such as neural networks and the internet
\citep{citeulike:99}.  The networks are characterized by having a
mixed structure \citep{Boccaletti06}, with connections between
geometrically nearby neighbors as well as long distance connections,
which drastically reduce the typical path length between any two
nodes. The HN3 network does not have the same mean-field properties
usually associated with such networks. For instance, average path
lengths scale with $\sqrt{L}$ instead of $\ln(L)$ with system size
$L$. But HN3 is constructed in a hierarchical manner that is conducive
for the renormalization group, which is a technique that takes
advantage of the self-similarity of systems \citep{SWN}.

\begin{figure}
\vskip 1.2in
\includegraphics{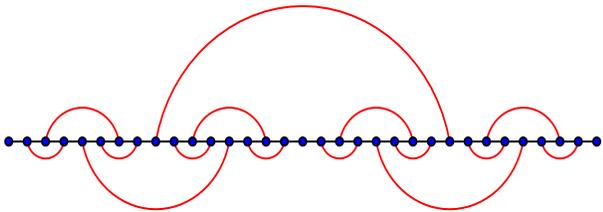}
\caption{\label{fig:hn3}
The HN3 network consists of a one-dimensional backbone
and a hierarchy of long distance connections.}
\end{figure}

HN3 consists of a one dimensional line as a backbone with $L=2^{k}+1$
sites, where $k$ is a positive integer which defines the number of
hierarchies. To make the long distance connections, we consider an
integer $i\le k$, which defines the level of the hierarchy, and $j$,
an integer which parameterizes the connection in a hierarchy. All
sites (except for $n=0$) are then uniquely represented by
\begin{equation}
n=2^{i}(2j+1).
\label{eq:hnparam}
\end{equation}
For example, for $i=0$, $n$ runs over all the odd integers, and
$i=1$ makes $n$ to be all integers once divisible by 2 (i.e. 2,
6, 10, ...), etc. Connections are made between neighbors within the
hierarchy, so for $i=0$, site 1 connects to 3, 5 to 7, etc, and for
$i=1$, 2 to 6, 10 to 14, etc. It is possible to make a more complex
network, HN4, by connecting two neighbors in the hierarchy to every
site. Fig.~\ref{fig:hn3} shows the HN3 network for $k=5$. The first,
middle, and last site are special in that they have no long range
connections (the first and last site, 0 and $L$, are not shown).
All the other sites are connected to three other sites. The distance
between the ends of the HN3 network are known to be proportional to
$\sqrt{L}$, which is similar to the diagonal on a square lattice
with $L$ sites.

In an implementation of TASEP on the HN3 network (which we shall refer
to as HN3-TASEP), on sites with a long-range \emph{forward} link,
particles have the possibility to move to two different sites. To
obtain an interesting dynamics we decided to have the chosen particle
attempt a jump to the long distance site first, and if that site is
occupied, to reach the short distance site as in $1d$-TASEP. For such a
site, representing half of all sites, jamming is somewhat alleviated,
as it can free itself with a higher probability. In turn, the other
half of all sites, possessing one extra \emph{incoming} link instead,
occupation and jamming is far more likely.

We will also study a family of models that interpolates between
$1d$-TASEP and HN3-TASEP, using a probabilistic update rule: At each
update, with a
probability $r$, a long-range jump is attempted first but if such
a jump does not succeed, a nearest-neighbor forward-jump is
attempted. For $r=0$, no long-range jump ever occurs, representing the
$1d$-TASEP case, while $r=1$ corresponds to the HN3-TASEP case. We
find that the nature of TASEP changes discontinuously for $r>0$, numerically
signaled by the disappearance of the MC phase (and the associated
$2nd$-order transitions), while the shock persists at any $r$. We suspect that
such discontinuity can be attributed to the broken particle-hole
symmetry, which only holds for strictly $r=0$.

\section{Numerical Results\label{sec:Numerical-Results}}

We implemented  a Monte Carlo simulation for HN3-TASEP 
as follows. First a particle is randomly chosen from $N$ particles
on the lattice plus one virtual particle. If a lattice particle is
selected, it is moved forward one site if that site is unoccupied.
If a virtual particle is selected, a random number between zero and
one is generated, and if it is less than $\alpha$, a particle is
placed at the first site of the lattice unless it is occupied. If
a particle is selected which is at the end of the lattice, a random
number is generated, and if it is less than $\beta$, that particle
is removed. A sequence of $N+1$ attempts to move particles constitute
one Monte Carlo Sweep (MCS). The number of lattice sites for every
result in this report was set at $L=1023$, unless otherwise stated.
The simulations were run for $10^{6}$ MCS, and the first $10^{5}$
MCS were discarded to allow the system reached a steady state. Current
and density for the system were recorded every 100 MCS and averaged
over to characterize the steady state.

\begin{figure}
\vskip 4.0in
\includegraphics{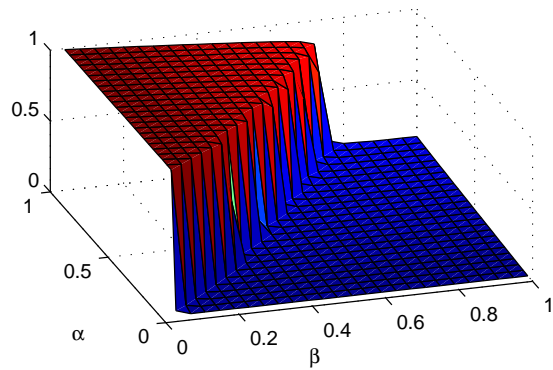}
\includegraphics{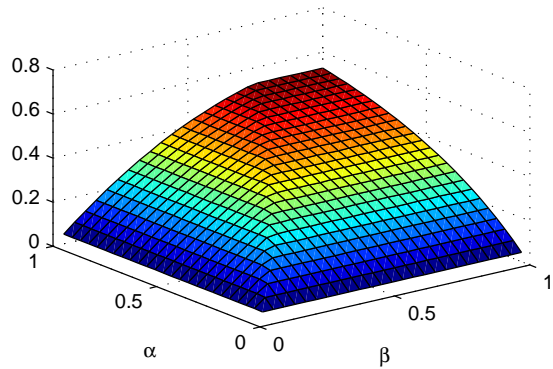}
\caption{\label{fig:tasep_hn3}to
compare with that of $1d$-TASEP in
Fig.~\ref{fig:Phasediagram-standard}
The density $\rho$ (top) and current $J$ (bottom) of TASEP on HN3 on a
grid of 25 values each for $\alpha$ and $\beta$. }
\end{figure}

\begin{figure}
\vskip 2.7in
\includegraphics{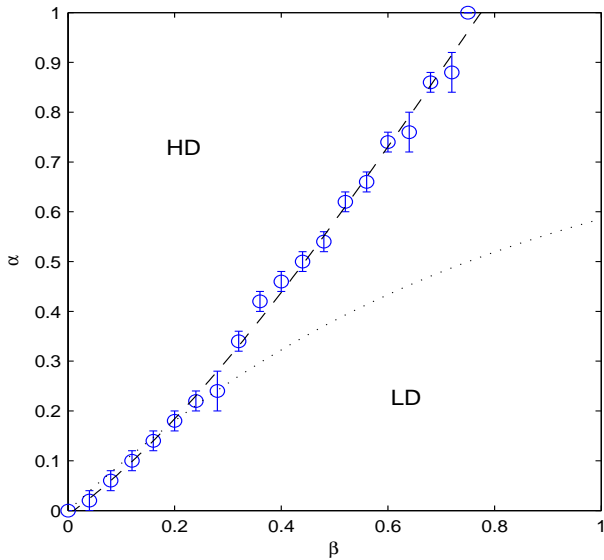}
\caption{\label{fig:Phasediagram-HN3}
Phase diagram of HN3-TASEP with only a high density (HD) and low density
(LD) phases, to
compare with that of $1d$-TASEP in
Fig.~\ref{fig:Phasediagram-standard}. The data points are extracted
from Fig.~\ref{fig:tasep_hn3}, the dashed line represents a simple
polynomial fit, and the dotted line is crude approximation to the
phase line resulting from the linearized mean-field equations in Sec.~\ref{Sec:Analytic}. }
\end{figure}

The behavior of HN3-TASEP is quite different from $1d$-TASEP
\citep{citeulike:1341557}, which can be seen when $\rho$ and $J$ are
plotted for a grid of 25 by 25 values of $\alpha$ and $\beta$ in
Fig.~\ref{fig:tasep_hn3}.  The HD and LD phases are present, but the
MC phase is conspicuously missing. The magnitude of $\rho$ and $J$ are
significantly altered. (Note that the current is measured here at the
exit site, which every particle moving through the system must pass;
the same is \emph{not} true for most other sites!) The density is much
higher in the HD phase and much lower in the LD phase when the HN3
shortcuts are present. In fact, throughout each phase, the density
remains almost constant.  The lattice is able to fill itself with
particles more efficiently in HD and remove them more efficiently in
LD when there are more connections between sites. Based on the density
plot in Fig.~\ref{fig:tasep_hn3}, in Fig.~\ref{fig:Phasediagram-HN3}
we extract the outlines of a phase diagram for HN3-TASEP to compare
with that of $1d$-TASEP in Fig.~\ref{fig:Phasediagram-standard}.

\begin{figure}
\vskip 4.6in
\includegraphics{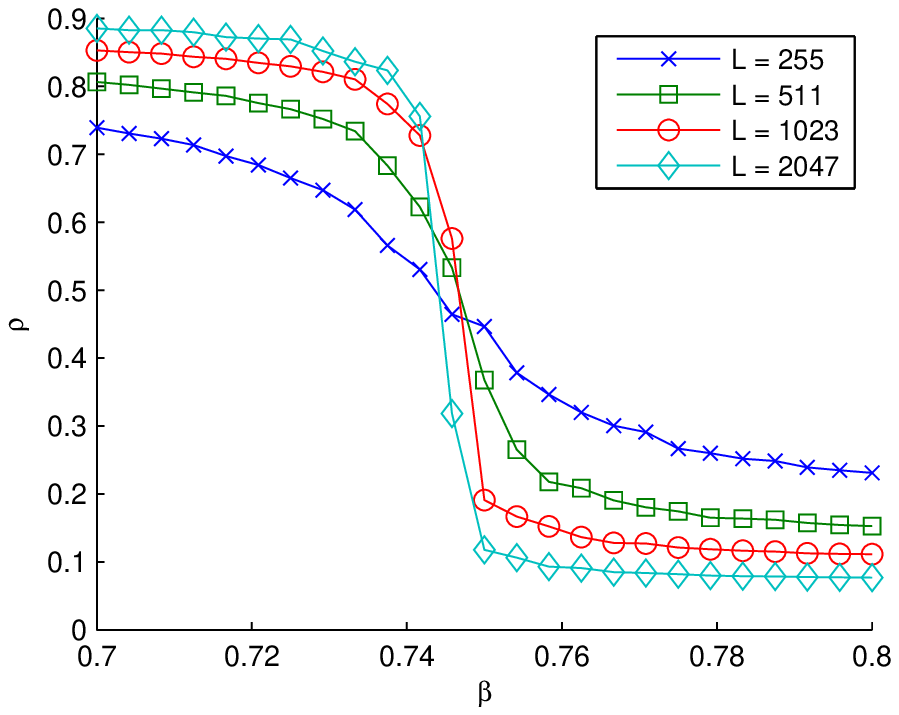}
\includegraphics{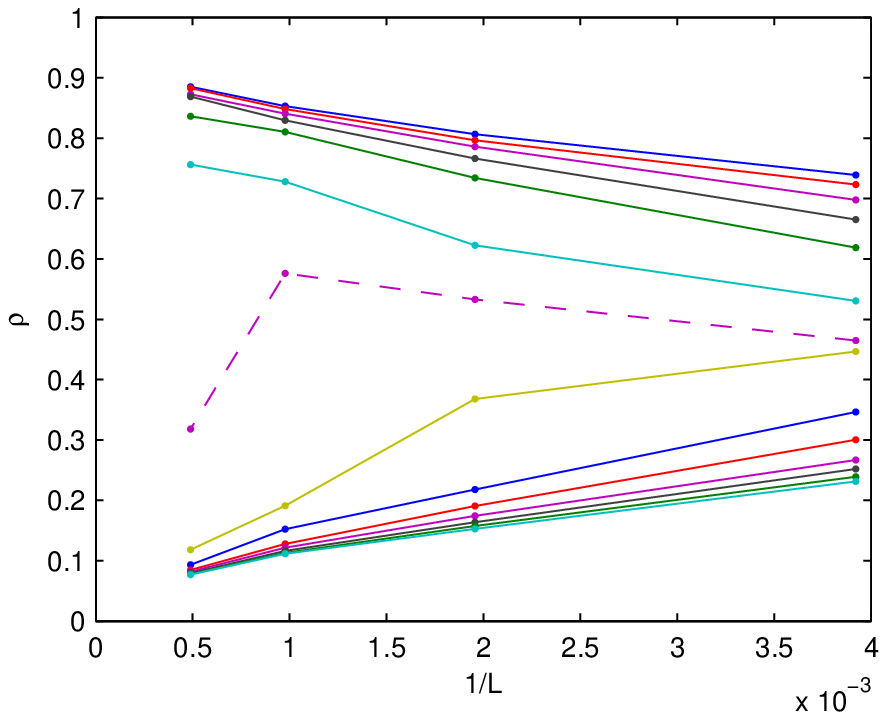}
\caption{\label{fig:transition} Top: Density during the transition from HD to
LD for lattices of size $L=255,$ 511, 1023, and 2047 at $\alpha=1$. The
total number of Monte Carlo Sweeps (MCS)  for each data point was
$5\,10^{6}$. Bottom: The density for the different values of $\beta$
plotted against $1/L$. The values range over $\beta=0.7\ldots0.8$
in steps of 0.08. The dashed line in the middle corresponds to $\beta=0.748$,
which appears closest to the transition, exhibits a significant error
as the stationary {}``shock'' state is hard to sample. But almost
all data below extrapolates to a value of $\rho$ consistent with
full packing, while all data for values of $\beta$ above the transition
extrapolates to an empty lattice, $\rho\approx0$. }
\end{figure}

Considering that HN3 is a hierarchical network, it is worth noting
that the results are smooth and not heterogeneous in any complicated
fashion (unlike the average occupation on sites, see below). Similar
to $1d$-TASEP, in the HD phase the current does not appear to vary
with $\alpha$, and in the LD phase it does not change appreciably with $\beta$. To
make sure the transition is first-order, $\rho$ is plotted for values
of $\alpha=1$ and $\beta$ between 0.7 and 0.8 for two lattice sizes in
Figs. \ref{fig:transition}.  The transition is less pronounced on a
smaller lattice, which indicates the transition is most likely sharp
in the thermodynamic limit: The data indicates that in the
thermodynamic limit $L\to\infty$ the lattice is completely filled
almost everywhere in the high-density phase, and it is completely
empty in the low-density phase. This result enables a simplifying Ansatz to
the equations in Sec.~\ref{Sec:Analytic} below.

\begin{figure}
\begin{centering}
\includegraphics{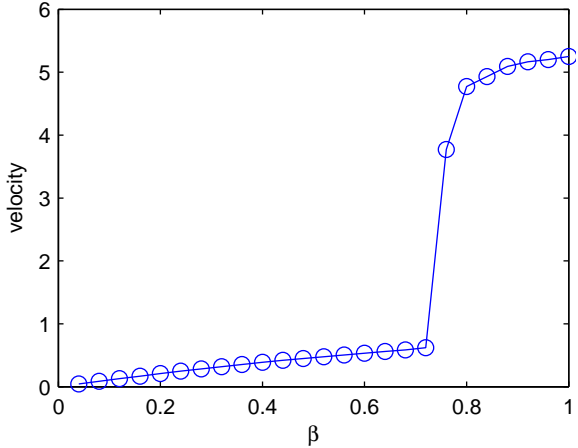}
\par\end{centering}
\caption{\label{fig:transition-velocity}Average velocity of particles for $\alpha=1$
and varying $\beta$. The velocity is $v<1$ in the HD phase ($\beta\lesssim0.75$),
and $v\gg1$ in the LD phase.}
\end{figure}

\begin{figure}
\begin{centering}
\includegraphics{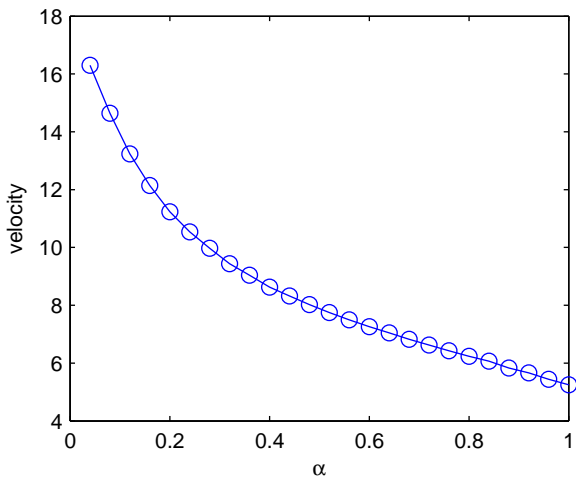}
\par\end{centering}
\caption{\label{fig:LD-velocity}Average velocity of TASEP on HN3 for $\beta=1$.}
\end{figure}

\begin{figure}
\begin{centering}
\includegraphics{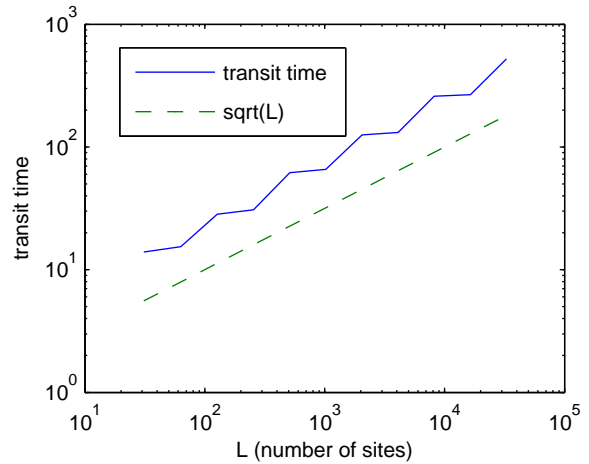}
\par\end{centering}
\caption{\label{fig:sqrtl}
Average transit time versus lattice size L for $\alpha=0.01$
and $\beta=1$. The transit time shows a $\sqrt{L}$ dependence, which
means particles are taking the shortest path on HN3. The stair-casing
is due to particles on networks with odd $k$ only taking the second shortest
path.}
\end{figure}

If one considers a single particle moving through an empty $1d$
lattice, it moves one site every step and $v=1$. Since in HN3 the
shortest end-to-end path is of length  $\sim\sqrt{L}$ \citep{SWPRL}, the time for a
particle to traverse it is $\sim\sqrt{L}$ and, hence,
$v=L/T\approx\sqrt{L}$.  The velocity is dependent on the lattice
size, and is unbounded for large lattices, although this definition of
velocity is based on the notion that the length of a lattice is equal
to it's number of sites. In the LD phase the particles can follow the
shortcuts, since they are usually unoccupied. In the HD phase most
sites, including ones at the end of shortcuts, are occupied, so the
particles have few opportunities to take a shortcut and they travel
along the backbone, which limits the velocity to
$v<1$. Fig.~\ref{fig:transition-velocity} shows the velocity for a
parameter range across the transition. In the HD phase the velocity is
indeed $<1$, but in the LD phase it is $\gg1$. The velocity can also
be examined for the case $\beta=1$, where the system is always in the
LD phase, as shown in Fig.~\ref{fig:LD-velocity}.  For small $\alpha$,
the number of particles is very low and they can take all of the
shortcuts. The velocity is on the order of $16\approx\sqrt{L}/2$.  As
$\alpha$ increases, some of the shortcuts are occupied and the
particles take trajectories that are a mix between shortcut and
backbone movements, which decreases the average velocity. The
$\sqrt{L}$-dependence of the transit time can be demonstrated through
finite size scaling.  Fig.~\ref{fig:sqrtl} shows the average transit
time at a low injection rate and high removal rate for many lattice
sizes, and it approximately follows $\sqrt{L}$. The stair-casing
effect between even and odd values of the hierarchy index $k$ is due
to the fact that HN3 is strictly self-similar only for every second
recursion of the hierarchy.  In particular, as particles move left to
right, they take every shortcut available, and on networks with odd
$k$ they miss the opportunity to take the largest shortcut, which is
part of the shortest path available.

\begin{figure}
\begin{centering}
\includegraphics{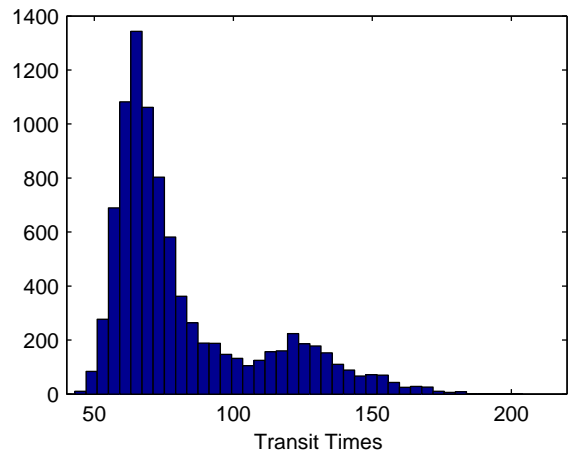}
\par\end{centering}
\caption{\label{fig:bimodal}Distribution of the transit times at $\alpha=0.1$,
and $\beta=1$.}
\end{figure}

One may also keep track of the number of time steps it takes for each
particle to cross the lattice. Fig.~\ref{fig:bimodal} shows a histogram
of the transit times for $\beta=$1 and a small value of $\alpha$,
so the particles are removed at the maximum rate and injected at a
slow rate. There is a bimodal distribution which suggests particles
are only taking a few distinct paths through the network, such as
the shortest and second shortest paths. It is interesting that the
bimodal distribution does not show up for other parts of the parameter
space. When $\alpha$ is smaller, the particles follow only the shortest
path, and there is only one peak. The width of the peak originates
with the update procedure, as not every particle is updated at every
Monte Carlo sweep, thus leading to fluctuations in transit times.
When $\alpha$ becomes larger there is some blockage of shortcuts
and the particles immediately follow many different paths. There is only one peak
in the distribution of transit times with a long tail and there is
no clear separation between fast moving particles and slow moving
particles. The number of particles taking the shortest path is probably
very small for any set of parameters except when the density is very
low.

\begin{figure}
\vskip 2.8in
\includegraphics{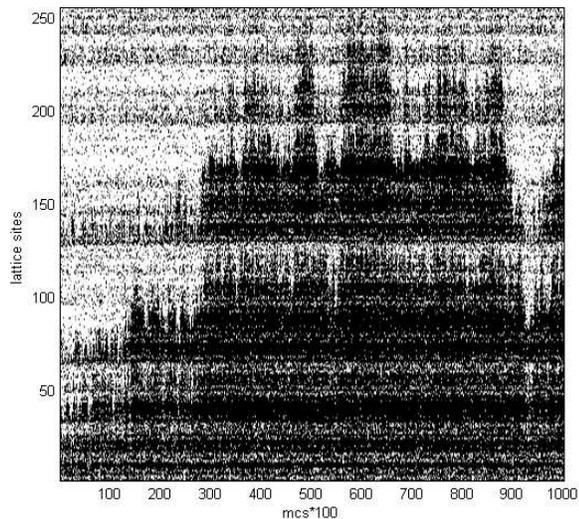}
\caption{\label{fig:shock-hn3}Time evolution of the shock phase in TASEP on
HN3 on a lattice of size $L=255$. Each white pixel is an occupied
site. Time is represented by 100 Monte Carlo sweeps.}
\end{figure}

We note that the shock phase looks very different on HN3 from that in
$1d$-TASEP. Fig.~\ref{fig:shock-hn3}
shows a typical time evolution of the occupation on the lattice at
a point in the parameter space where $\rho$ was roughly 0.5 and can
be compared to the shock phase in $1d$-TASEP \citep{Hinrichsen00}.
It is clear that there is coexistence of the LD and HD phases, but
there appears to be a hierarchy of boundary sites, depending on the
range of the long connections of the originating site, as we might 
expect that there is more movement near sites with the longer connection.

\begin{figure}
\vskip 5.4in
\includegraphics{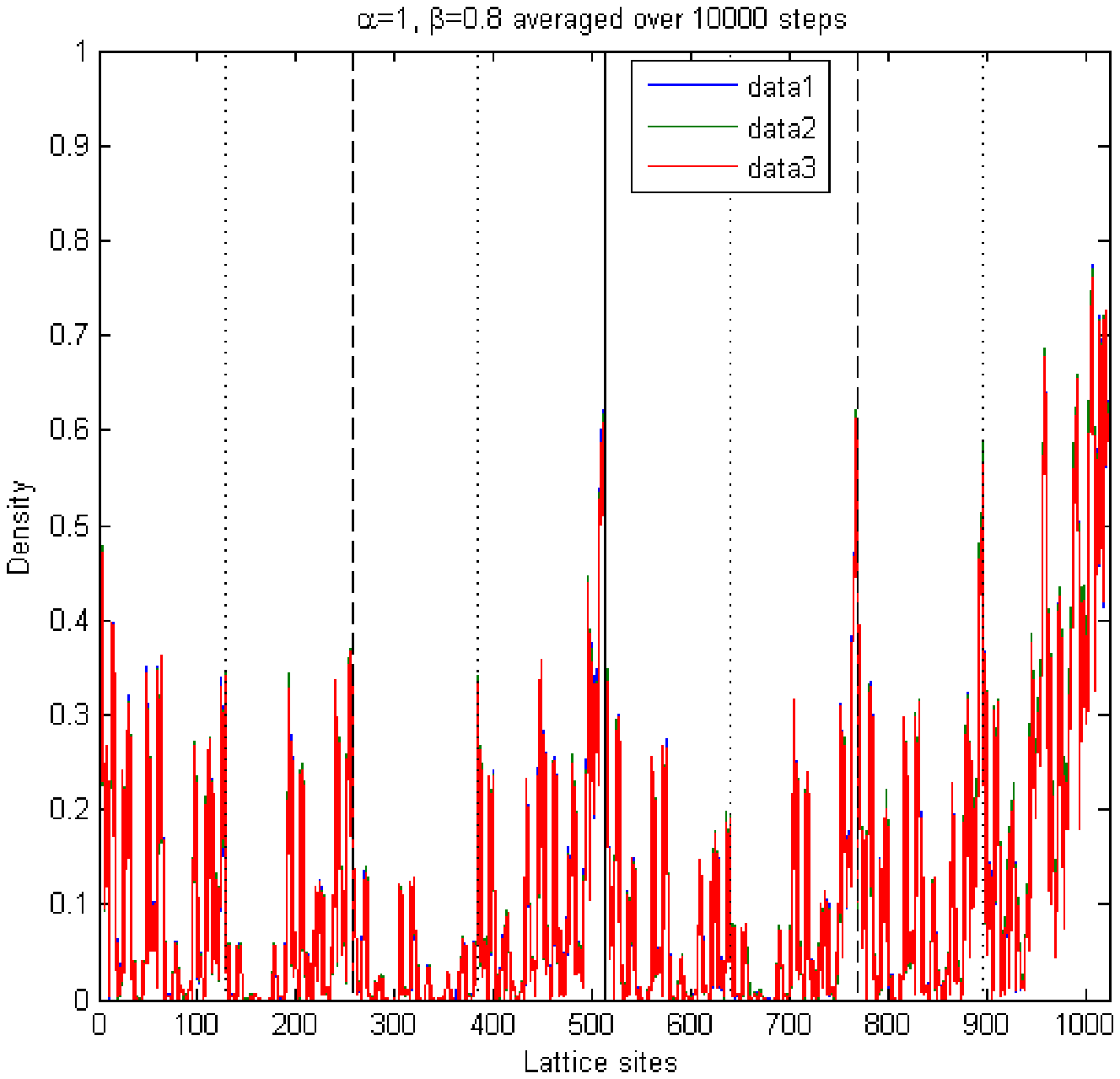}
\includegraphics{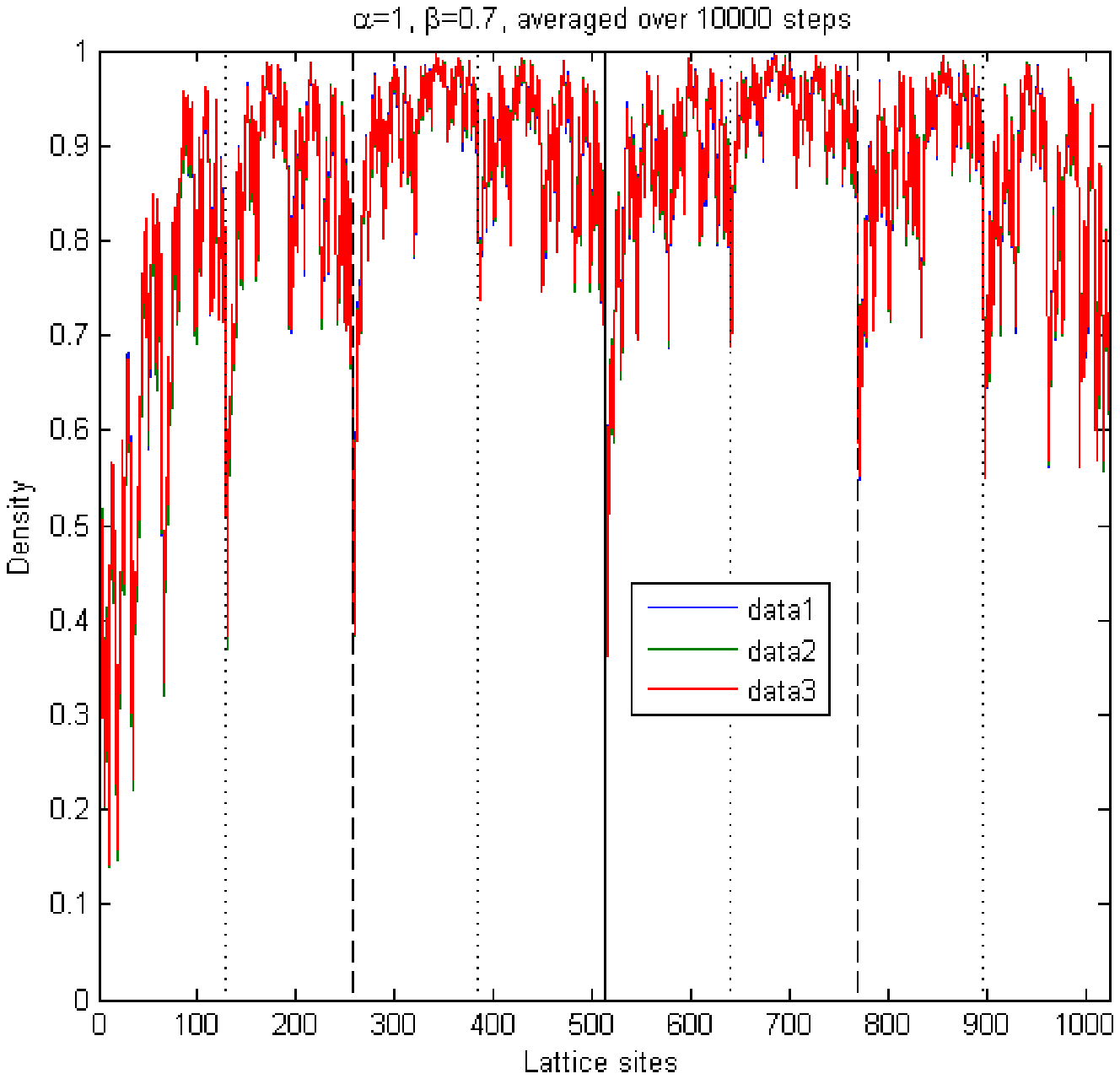}
\caption{\label{fig:occupation-hn3}
Average occupation $\left\langle \tau_{i}\right\rangle $
in the steady state on sites along the lattice backbone in HN3-TASEP
in LD (top) and HD (bottom). In low density (LD), sites with the longer-range
connections show more occupation as way-stations of short paths through
the lattice that bypass particles away from sites belonging to lower
levels of the hierarchy. In turn, in high density (HD) low-level sites
get jammed up while sites with longer-range connections (especially
forward) have an easier time to empty themselves. Note the boundary
layers of increased jamming near the exit in  LD and that of depletion
at the entrance in HD.}
\end{figure}

A similar picture about the intricate internal dynamics within the
lattice emerges from the average steady-state occupation $\left\langle
\tau_{i}\right\rangle $ for each site, shown in
Fig.~\ref{fig:occupation-hn3}. Smoothed over many sites, one could
argue that the general trend is similar to that in $1d$-TASEP, where
in LD the bulk density is constant, cumulating in a defined boundary
layer near the exit, while in HD there is such a layer at the
entrance, followed by a constant density plateau in the bulk
\citep{citeulike:1341557}. Although in HN3-TASEP corresponding
boundary layers are visible, the site-to-site density is extremely
rough, owing to the heterogeneous mix of incoming and outgoing
long-range jumps that belong to very different levels of the hierarchy
and thus have very different efficiencies in transmitting
particles. Our linear approximation to the steady-state equations in
Sec.~\ref{Sec:Analytic} manages to reproduce these heterogeneities
very well.

\begin{figure}
\vskip 2.2in
\includegraphics{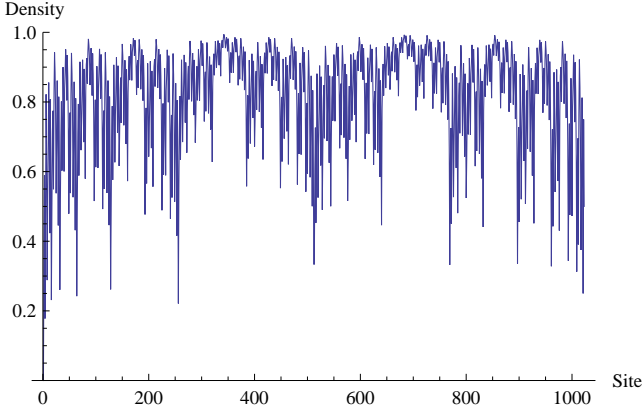}
\caption{\label{fig:DensityApprox-hn3}
Average steady-state occupation $\left\langle
\tau_{i}\right\rangle\approx1-\epsilon_i$ on HN3-TASEP, as obtained analytically
in the linearized approximation ($\epsilon_i\ll1$) in
Eq.~(\ref{eq:LinearApprox}) for HD. Aside from an arbitrary overall scale
for the $\epsilon_i$, the pattern corresponds in great detail to
that in Fig.~\ref{fig:occupation-hn3}, except for the boundaries.}
\end{figure}

\section{Analytic Treatment}\label{Sec:Analytic}
To obtain a set of master equations such as those in
Eq.~(\ref{eq:master2}) for HN3-TASEP, we have
to distinguish two different kinds of sites. An IN-site $i$
simply has a long-range connection to a site $i-l$, preceding it by a
distance $l$ in the lattice, in addition to its immediate predecessor
and successor sites $i-1$ and $i+1$. Generically, at any level in the
hierarchy, updates at an IN-site $i$ do not affect either of the
predecessor sites, hence, the equations are similar to those in
Eqs.~(\ref{eq:master2}):
\begin{eqnarray}
\tau_{i}(t+dt)&=&\tau_{i}(t)\tau_{i+1}(t),\label{eq:INi}\\
\tau_{i+1}(t+dt)&=&\tau_{i+1}(t)+\left[1-\tau_{i+1}(t)\right]\tau_{i}(t).
\nonumber
\end{eqnarray}
In turn, when the update occurs at an OUT-site $i$ with a long-range
link to a successor site $i+l$, that site $i$, the immediate successor
site $i+1$, \emph{and} the long-range successor site $i+l$ are
affected in a novel way in HN3-TASEP. In our choice of moving
particles preferentially along the long links leads to a three-stage
update process: Only if the long jump is blocked, a short jump to the
successor site is attempted; if that is blocked as well, the particles
remains on its site. These choices are expressed through the following
equations for the updated site, its successor, and its long-range
forward neighbor:
\begin{eqnarray}
\tau_{i}(t+dt) & = & \tau_{i}(t)\tau_{i+1}(t)\tau_{i+l}(t),\label{eq:OUTi}\\
\tau_{i+1}(t+dt) & = & \tau_{i+1}(t)+\tau_{i}(t)\left[1-\tau_{i+1}(t)\right]\tau_{i+l}(t),\nonumber\\
\tau_{i+l}(t+dt) & = &
\tau_{i+l}(t)+\tau_{i}(t)\left[1-\tau_{i+l}(t)\right].\nonumber
\end{eqnarray}
Note that these equations are inherently cubic in the dynamic
variables.  Another complication is the lack of translational
invariance, as these equations depend on a forward-distance $l$, which
itself depends strongly on the hierarchical level that site $i$
belongs to. At least, the boundary conditions, affecting sites $i=0$
and $i=L$, are identical to $1d$-TASEP.

To make any progress at all, we already at this point consider the
mean-field limit. The mean-field limit averages the dynamic variables
over the noise, eliminating fluctuations and correlations, i.~e. we
set $\left\langle \tau_{i}\tau_{j}\right\rangle \sim\left\langle
\tau_{i}\right\rangle \left\langle \tau_{j}\right\rangle
=\tau_{i}\tau_{j}$, and allows us to arrive at a set of rate equations
for the continuous variables $\tau_{i}$. We have to take full account
of the lack of translational invariance along the line in HN3 by
addressing the hierarchical level any site belongs to. All sites on
the lowest level are of odd index $2j+1$, and they are alternately IN
and OUT-sites. Say, all sites $4j+1$ are OUT-sites. While it is then
clear that it has two successors, one and two steps ahead, and a
single predecessor site, the latter itself may be an IN or an
OUT-site. But depending on that, its update will affect our $4j+1$
site differently, and we make the further simplifying assumption of an
equal balance between both possibilities. Hence, on average, site
$4j+1$ is changed as
\begin{eqnarray}
\tau_{4j+1}(t+dt)&=&\left(1-dt\right)\tau_{4j+1}(t)\label{eq:OUT4j} \\
&& +dt\,\tau_{4j+1}(t)\tau_{4j+2}(t)\tau_{4j+3}(t)\nonumber\\
 &  & +\frac{dt}{2}\,\tau_{4j}(t)\left[1-\tau_{4j+1}(t)\right]\nonumber\\
 &  &+\frac{dt}{2}\,\tau_{4j}(t)\tau_{4j+l}(t)\left[1-\tau_{4j+1}(t)\right],\nonumber
\end{eqnarray}
where the distance $l$ furthermore depends on that predecessor site at
$4j$. In order, the terms in Eq.~(\ref{eq:OUT4j}) refer to either
nothing changing with probability $1-dt$, our reference site $4j+1$
being updated itself {[}see the first of Eqs.~(\ref{eq:OUTi})] with
probability $dt$, or the predecessor site $4j$ being update with
probability $dt/2$ for each of the two scenarios given. Similar
considerations holds for the IN-sites at $4j+3$, with yet another
complication arising from the incoming long-range predecessor site
$4j+1$, whose potential update adds another term:
\begin{eqnarray}
\tau_{4j+3}(t+dt) &=& \left(1-dt\right)\tau_{4j+3}(t)\label{eq:IN4j} \\
&& +dt\,\tau_{4j+3}(t)\tau_{4j+4}(t)\nonumber\\
&& +\frac{dt}{2}\,\tau_{4j+2}(t)\left[1-\tau_{4j+3}(t)\right]\nonumber\\
&& +\frac{dt}{2}\,\tau_{4j+2}(t)\tau_{4j+2+l}(t)\left[1-\tau_{4j+3}(t)\right] \nonumber\\
&& +dt\,\tau_{4j+1}(t)\left[1-\tau_{4j+3}(t)\right].\nonumber
\end{eqnarray}
At all other levels in the hierarchy, matters somewhat simplify, as
\emph{any} even site clearly has an odd-indexed site preceding and
following it, making the effect of long-range bonds fully apparent.
In particular, sites $2(2j+1)$ at the next-to-bottom level \emph{all}
have an odd OUT-site $4j+1$ preceding it and the corresponding IN-site
$4j+3$ following it, independent of whether they themselves are OUT or
IN-sites. Still, in their own update behavior, OUT-sites at $2(4j+1)$
and IN-sites at $2(4j+3)$ at this level differ, and we get
\begin{eqnarray}
&&\tau_{2\left(4j+1\right)}(t+dt)=\left(1-dt\right)\tau_{2\left(4j+1\right)}(t)\label{eq:OUT8j}\\
&&\quad+dt\,\tau_{2\left(4j+1\right)}(t)\tau_{2\left(4j+1\right)+1}(t)\tau_{2\left(4j+3\right)}(t) \nonumber\\
&&\quad +dt\,\tau_{2\left(4j+1\right)-1}(t)\tau_{2\left(4j+l\right)+1}(t)\left[1-\tau_{2\left(4j+1\right)}(t)\right], \nonumber
\end{eqnarray}
and
\begin{eqnarray}
&&\tau_{2\left(4j+3\right)}(t+dt) =
  \left(1-dt\right)\tau_{2\left(4j+3\right)}(t)\label{eq:IN8j}\\
&&\quad+dt\,\tau_{2\left(4j+3\right)}(t)\tau_{2\left(4j+3\right)+1}(t) \nonumber\\
&&\quad +dt\,\tau_{2\left(4j+3\right)-1}(t)\tau_{2\left(4j+3\right)+1}(t)\left[1-\tau_{2\left(4j+3\right)}(t)\right]\nonumber\\
&&\quad +dt\,\tau_{2\left(4j+1\right)}(t)\left[1-\tau_{2\left(4j+3\right)}(t)\right]. \nonumber
\end{eqnarray}
Finally, at any higher level $i\geq2$ of the hierarchy, all sites
$2^{i}\left(2j+1\right)$ only have odd-indexed IN-sites as predecessor
and odd-indexed OUT-sites as successor. We get for OUT and IN-sites,
respectively:
\begin{eqnarray}
&&\tau_{2^{i}\left(4j+1\right)}(t+dt)=
  \left(1-dt\right)\tau_{2^{i}\left(4j+1\right)}(t)\label{eq:OUTalli}\\
&&\qquad+dt\,\tau_{2^{i}\left(4j+1\right)}(t)\tau_{2^{i}\left(4j+1\right)+1}(t)\tau_{2^{i}\left(4j+3\right)}(t) \nonumber\\
&&\qquad +dt\,\tau_{2^{i}\left(4j+1\right)-1}(t)\left[1-\tau_{2^{i}\left(4j+1\right)}(t)\right],  \nonumber
\end{eqnarray}
and
\begin{eqnarray}
&&\tau_{2^{i}\left(4j+3\right)}(t+dt)=
  \left(1-dt\right)\tau_{2^{i}\left(4j+3\right)}(t)\label{eq:INalli}\\
&&\qquad+dt\,\tau_{2^{i}\left(4j+3\right)}(t)\tau_{2^{i}\left(4j+3\right)+1}(t) \nonumber\\
&&\qquad+dt\,\tau_{2^{i}\left(4j+3\right)-1}(t)\left[1-\tau_{2^{i}\left(4j+3\right)}(t)\right] \nonumber\\
&&\qquad+dt\,\tau_{2^{i}\left(4j+1\right)}(t)\left[1-\tau_{2^{i}\left(4j+3\right)}(t)\right].\nonumber
\end{eqnarray}

In the steady state, we consider $t\sim t+dt\to\infty$. Then, the
above equations somewhat simplify, and we get from
Eqs.~(\ref{eq:OUT4j}-\ref{eq:IN4j}):
\begin{eqnarray}
\tau_{n} & = & \tau_{n}\tau_{n+1}\tau_{n+2}+\frac{1}{2}\tau_{n-1}\left(1-\tau_{n}\right)\left(1+\tau_{n-1+l}\right),\quad\label{eq:level0mf}\\
\tau_{m} & = & \tau_{m}\tau_{m+1}+\frac{1}{2}\tau_{m-1}\left(1-\tau_{m}\right)\left(1+\tau_{m-1+l}\right) \nonumber\\
&&\qquad+\tau_{m-2}\left(1-\tau_{m}\right),\nonumber
\end{eqnarray}
at the lowest level, where we have set $n=4j+1$ for OUT-sites and
$m=4j+3$ for IN-sites. Similarly, from Eqs.~(\ref{eq:OUT8j}-\ref{eq:IN8j}),
we get
\begin{eqnarray}
\tau_{n} & = & \tau_{n}\tau_{n+1}\tau_{n+4}+\tau_{n-1}\left(1-\tau_{n}\right)\tau_{n+1},\label{eq:level1mf}\\
\tau_{m} & = & \tau_{m}\tau_{m+1}+\tau_{m-1}\left(1-\tau_{m}\right)\tau_{m+1} \nonumber\\
&&\qquad+\tau_{m-4}\left(1-\tau_{m}\right),\nonumber
\end{eqnarray}
here setting $n=2(4j+1)$ for the OUT-sites and $m=2(4j+3)$ for the
IN-sites. Finally, the general case in the hierarchy from Eqs.~(\ref{eq:OUTalli}-\ref{eq:INalli})
reduces to
\begin{eqnarray}
\tau_{n} & = & \tau_{n}\tau_{n+1}\tau_{n+2^{i+1}}+\tau_{n-1}\left(1-\tau_{n}\right),\label{eq:level1ALLmf}\\
\tau_{m} & = & \tau_{m}\tau_{m+1}+\tau_{m-1}\left(1-\tau_{m}\right) \nonumber\\
&&\qquad+\tau_{m-2^{i+1}}\left(1-\tau_{m}\right),\nonumber
\end{eqnarray}
with $n=2^{i}(4j+1)$ for the OUT-sites and $m=2^{i}(4j+3)$ for the
IN-sites for $i\geq2$. Since the first and last bond in the lattice do
not get bridged by a long-range jump, see Fig.~\ref{fig:hn3}, the
boundary conditions are similar to those for $1d$-TASEP
\citep{citeulike:1341557}:
\begin{eqnarray}
\tau_{0}\left(1-\tau_1\right) & = & \alpha\left(1- \tau_{0}\right),\label{eq:BC}\\
\beta\tau_{L} & = & \tau_{L-1}\left(1-\tau_{L}\right),\nonumber
\end{eqnarray}
with $L=2^k$.

Although the steady state has simplified the equations somewhat, their
main difficulties remain their lack of symmetry and their cubic order.
Yet, our numerical results in Sec.~\ref{sec:Numerical-Results},
particularly Fig.~\ref{fig:tasep_hn3}, suggest a very simple phase
structure with $\tau_{i}\sim1-\epsilon(\alpha,\beta)$ in the bulk
throughout HD, and $\tau_{i}\sim\delta(\alpha,\beta)$ throughout LD,
with $\epsilon,\delta\ll1$. Already a linear approximation in HD
provides some illuminating insight.\footnote{Unfortunately, a mere
  linear approximation is insufficient in LD} To first order in
$\epsilon_i$, the distinctions between the different orders in the
hierarchy, as expressed in
Eqs.~(\ref{eq:level0mf}-\ref{eq:level1ALLmf}), disappear to leave for
all $0\leq i\leq k-2$ and $0\leq j\leq2^{k-2-i}$:
\begin{eqnarray}
\epsilon_{2^{i}\left(4j+1\right)}
&\sim&\epsilon_{2^{i}\left(4j+1\right)+1}+\epsilon_{2^{i}\left(4j+3\right)}, \nonumber\\
\epsilon_{2^{i}\left(4j+3\right)}&\sim&\frac{1}{2}\epsilon_{2^{i}\left(4j+3\right)+1}.
\label{eq:LinearApprox}
\end{eqnarray}
Special rules apply for the boundary sites in Eqs.~(\ref{eq:BC}):
\begin{eqnarray}
\left(1-\epsilon_{0}\right)\epsilon_1 &=& \alpha\epsilon_0, \nonumber\\
\beta\left(1-\epsilon_{L}\right)&=&\left(1-\epsilon_{L-1}\right)\epsilon_{L},
\label{eq:LinearApproxBC}
\end{eqnarray}
which we leave unexpanded for now,  and the central site:
\begin{eqnarray}
\epsilon_{2^{k-1}} &\sim& \epsilon_{2^{k-1}-1},
\label{eq:LinearApproxCenter}
\end{eqnarray}
each of which lacks a long-range bond. The solution for all $\epsilon_i$, while not
presentable in closed form, is easily obtained in $O(L)$ steps in terms of
$\epsilon_L$, starting from the exit
backwards. Fig.~\ref{fig:DensityApprox-hn3} demonstrates the quality
of this approximation in accounting in great detail for the Weierstrass-like
hierarchical nature \citep{SWN} of the site densities in  HD found in
Fig.~\ref{fig:occupation-hn3}. In particular, the pattern on bulk
sites is well represented, aside from an overall scale represented by
the value of $\epsilon_L$ that depends on $\alpha$ and
$\beta$. Naturally, sites near the boundaries are less-well
approximated, since the $\epsilon_i$ are not sufficiently small
there. 

Current conservation
on the first and last bond dictates
\begin{eqnarray}
\epsilon_1 &\sim& \epsilon_L,
\label{eq:1L}
\end{eqnarray}  
which is \emph{automatically} satisfied by
Eqs.~(\ref{eq:LinearApprox}). Together with
Eq.~(\ref{eq:LinearApproxBC}), Eq.~(\ref{eq:1L}) relates the boundary
conditions together, which allows to fix the arbitrary scale in terms
of  $\alpha$ and
$\beta$. More interesting, we can estimate the critical line
$(\alpha_c,\beta_c)$ between HD and LD as the location where the
linear approximation breaks down. Eliminating the other unknowns, we
find the nontrivial relation involving both boundary conditions:
\begin{eqnarray}
\beta&=&\frac{\alpha\epsilon_0}{1-\epsilon_0}\frac{1-\frac{1}{2}\frac{\alpha\epsilon_0}{1-\epsilon_0}}{1-\frac{\alpha\epsilon_0}{1-\epsilon_0}},
\label{eq:Crit_e0}
\end{eqnarray}
that fixes $\epsilon_0(\alpha,\beta)$. In turn, any such relation is
only reasonable for sufficiently small $\epsilon_0$. Since
$\epsilon_0$ varies between 0 and 1, either extreme providing a
trivial or singular result, respectively, we chose (somewhat arbitrarily) a generic
in-between value of  $\epsilon_0=1/2$ to estimate the phase line. The
resulting relation, $\alpha=1+\beta-\sqrt{1+\beta^2}$, is also indicated
in the phase diagram in Fig.~\ref{fig:Phasediagram-HN3}.

\begin{figure}
\begin{centering}
\includegraphics{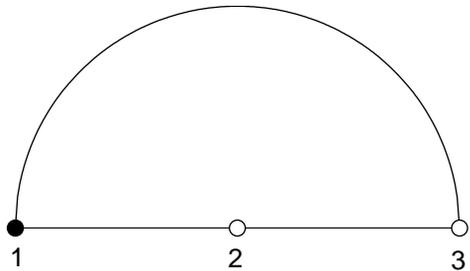}
\par\end{centering}
\caption{\label{fig:part-hole}
Particle-hole symmetry is broken by the particle's preference to take
the shortcut. A particle may move from site 1 to site 3, but a hole
may not move from site 2 to site 1.}
\end{figure}

\begin{figure}
\vskip 2.5in
\includegraphics{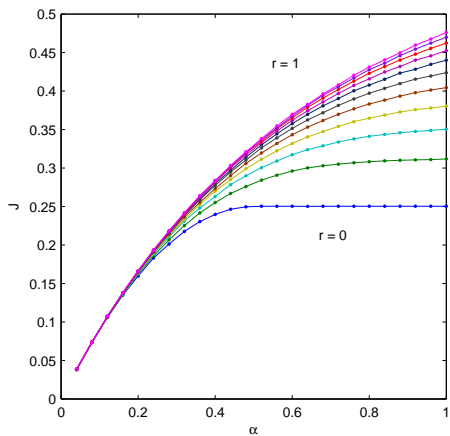}
\caption{\label{fig:rfamilycurrent}
Plot of the current $J(\alpha)$ for
  fixed $\beta=1$ in the one-parameter family of models described in
  the text for $0\leq r\leq1$ in a system of size $L=1024$. For $r=0$ only, the case of $1d$-TASEP,
  particle-hole symmetry
  is obeyed and phase transition into an MC phase occurs. For all
  $r>0$, the symmetry is broken and $J(\alpha)$ varies smoothly,
  similar to $r=1$, the case of HN3-TASEP.
}
\end{figure}

\section{Discussion}
\label{sec:Discussion}
We found that there are many striking differences between $1d$-TASEP and TASEP on
HN3. It is obvious that particle-hole symmetry is broken from the
asymmetric phase diagram, where the phase boundary between HD and LD
curves and favors a larger LD phase. The symmetry is broken due to the
particle's preference to take shortcuts, and Fig.~\ref{fig:part-hole}
illustrates the most obvious case of this symmetry breaking. A
particle may only move from site 1 to site 3, but a hole may not move
from site 2 to site 1 since the particle may not move from site 1 to
site 2. A full lattice with a single hole moving to the left cannot
take every shortcut as does a single particle on an empty lattice
moving to the right. The model favors the efficient flow of particles,
and this is why the LD phase is larger and free flow persists for
some values of $\alpha>\beta$.

\begin{figure}
\vskip 5.4in
\includegraphics{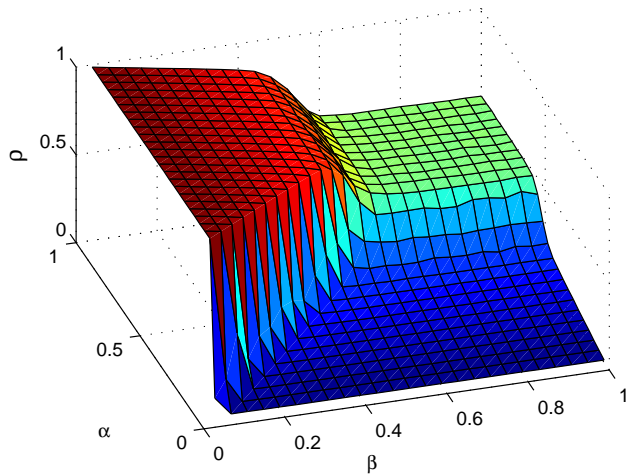}
\includegraphics{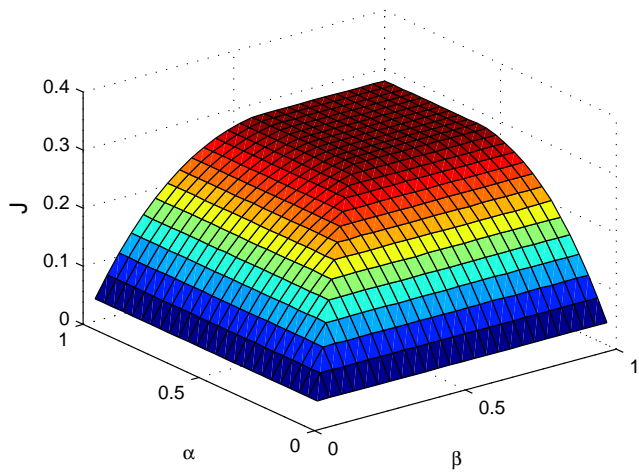}
\caption{\label{fig:phsym}
Density $\rho$ (top) and current $J$ (bottom) of TASEP on HN3 with
particles having 50\% chance of only attempting long distance jump and
50\% chance of only attempting jumps along the backbone.}
\end{figure}

The importance of this symmetry is highlighted by designing a one-parameter family
of models that interprets between  $1d$-TASEP and HN3-TASEP. In these
models, particles attempt a long-range jump with probability $r$,
$0\leq r\leq1$. Should no jump result, either because no attempt was
made or because the attempt failed, a nearest-neighbor jump is
attempted. For $r=0$, only nearest-neighbor jumps occur, corresponding
to $1d$-TASEP. In turn, for $r=1$, a long-range jump is always
attempted first, as in HN3-TASEP. We can characterize the behavior of
this model with $r$ sufficiently by presenting our numerical results
for the current $J$ as a function of $\alpha$ for fixed $\beta=1$.
Fig.~\ref{fig:rfamilycurrent} shows how the smooth variation of
$J(\alpha)$ found for HN3-TASEP in Fig.~\ref{fig:tasep_hn3} evolves
from $r=1$ towards the $2nd$-order phase transition between LD and MC
phases at $r=0$. Although finite-size effects (at $L=1024$ here)
obscure the bend in $J$ at $\alpha=1/2$, it is quite clear that only
for strictly $r=0$ there is a MC phase with a constant-current
plateau. For any $r>0$, $J(\alpha)$ appears to remain a smooth
function without the emergence of a plateau. Of course, for \emph{all} such
$r>0$, particle-hole symmetry is broken.

We can formulate an update rule on HN3 that \emph{does} preserve
particle-hole symmetry. For instance, the particle chosen for an
update could attempt a long-range or a nearest-neighbor jump with
probability $r$ and $1-r$, respectively, but \emph{not} explore the
alternative if such an attempt fails. Unfortunately, although $r=0$
again corresponds to $1d$-TASEP, $r=1$ does not attain HN3-TASEP but another
version of $1d$-TASEP restricted to $\sim\sqrt{L}$ sites only. In
Fig.~\ref{fig:phsym} we display the density and the current as a
function of $\alpha$ and $\beta$ for the case $r=1/2$. The symmetry
with respect to $\alpha=\beta$ is easily visible. As in $1d$-TASEP,
there are HD, LD, and MC phases with a shock line between HD and
LD. The most notable differences are that $J>1/4$ in MC and the steep
transitions in the density at each phase boundary.

Returning to HN3-TASEP, we found in the LD phase that the particles
move freely across the lattice. As the injection rate $\alpha$ is
increased, the density and current increase, and there are few
collisions, but the network is still able to transport particles
quickly. At the transition into the HD phase, jams snowball throughout
the system and soon fill in the whole lattice. The shortcuts are
likely to be blocked off and the particles slowly crawl along the
backbone. One may make an analogy to vehicular traffic. Cars like
particles can move along a one lane road, and they may not pass each
other. They appear at the beginning of the road at a rate $\alpha$,
and disappear at the end of the road at a rate $\beta.$ At some
density of cars the average velocity drops significantly and the whole
road is jammed. If we add highways then cars may enter a highway, go
faster, and when they get off they end up in front of their peers. In
free flowing traffic, highways reduce the amount of time for cars to
reach their destination and lower the density, given the same boundary
conditions.  Traffic jams still occur with the addition of highways,
and there are more possibilities for cars to maneuver around each
other and fill in gaps so the density can be very high.

An interesting property of TASEP on HN3 is that the network uses
mainly it's long range connections to transport particles, and its
backbone to 'store' them. In the LD phase the particles flow freely
through the highway. As $\alpha$ is increased, particles may be pushed
into the less visited backbone sites, and the current and density
increase.  An inverse process happens in the HD phase, where as
$\beta$ is increased more particles can be pulled from storage, the
current goes up, and the density goes down. There is no possibility
for the current and density to stay constant as both parameters are
changed simultaneously, hence, there is no MC phase on HN3.

This storage mechanism could conceivably be useful in a real system.
The transition between the HD and LD phases changes the system from
very high density to very low density, while the current does not
change significantly. Only a small local change in the rate of
particle injection or removal at the boundaries can induce a dramatic
global change in the number of particles.  We have measure the effect
of a protocol whereby at $\alpha=1$ the value of $\beta$ is switched
between 0.7 (HD phase) and 0.8 (LD phase), see also
Fig.~\ref{fig:transition}.  Unfortunately, it takes a long time, about
a factor of 1000 longer than the typical transit time of a particle,
to squeeze the excess particles through the single exit site, empty
out the lattice and re-establish the steady state at $\beta=0.8$. In
turn, jamming up the system by re-setting to $\beta=0.7$ attains it
steady state at least an order of magnitude faster.

\section{Conclusions}
\label{sec:Conclusions}
In this work we found that a variation of the TASEP with fixed long
distance jumps can lead to significant changes in its phases, notably,
the disappearance of the maximum current phase. This is surprising
since the TASEP phases are considered to be robust to many changes.
These changes were described qualitatively as a result of the
particles velocity becoming essentially unbounded along paths
utilizing the long jumps.

Despite the increased complexity of the rate equations for this model,
the much-simplified phase structure we found allowed the possibility
of an analytic approach that, even if approximate, should provide a
good qualitative and quantitative description. While a more detailed
solution eludes us here, it remains a worthwhile goal for the future,
as it would provide novel insight into and control over a process that
has stimulated significant advances in the understanding of
non-equilibrium critical phenomena.

\bibliographystyle{unsrt}
\bibliography{/Users/stb/Boettcher}

\begin{thebibliography}{10}

\bibitem{Schmittmann95}
B.~Schmittmann and R.~K.~P. Zia in
\newblock{\em Phase Transitions and Critical Phenomena, Vol. 17.}
\newblock (Academic Press, London), 1995.

\bibitem{citeulike:1532138}
C.~T. MacDonald, J.~H. Gibbs, and A.~C. Pipkin.
\newblock {\em Biopolymers}, 6:1, 1968.

\bibitem{Barabasi95}
A.-L. Barabasi and H.~E. Stanley.
\newblock {\em Fractal Concepts in Surface Growth}.
\newblock Cambridge University Press, 1995.

\bibitem{Biham92}
O. Biham, A.~A. Middleton, and D. Levine.
\newblock {\em Phys. Rev. A}, 46:R6124, 1992.

\bibitem{Nagel92}
K.~Nagel and M.~Schreckenberg.
\newblock {\em J. Phys. I France}, 2:2221, 1992.

\bibitem{Nagel95}
K. Nagel and M. Paczuski.
\newblock {\em Phys. Rev. E}, 51:2909, 1995.

\bibitem{citeulike:3175269}
O.~G. Berg, R.~B. Winter, and P.~H. von Hippel.
\newblock {\em Biochemistry}, 20:6929, 1981.

\bibitem{citeulike:1341557}
B.~Derrida, E.~Domany, and D.~Mukamel.
\newblock {\em J. Stat. Phys.}, 69:667, 1992.

\bibitem{citeulike:3196043}
B.~Derrida, M.~R. Evans, V.~Hakim, and V.~Pasquier.
\newblock {\em J. Phys. A: Math. Gen.}, 1493, 1993.

\bibitem{citeulike:3196051}
G.~Sch\"{u}tz and E.~Domany.
\newblock {\em  J. Stat. Phys.}, 72:277, 1993.

\bibitem{Hinrichsen00}
H.~Hinrichsen.
\newblock {\em Advances in Physics}, 49:815, 2000.

\bibitem{Mobilia08}
M. Mobilia, T. Reichenbach, H. Hinsch, T. Franosch, and E.  Frey.
\newblock {\em Banach Center Publications}, 80:101, 2008.

\bibitem{citeulike:3163805}
J. Szavits-Nossan and K.~Uzelac.
\newblock {\em Phys. Rev. E}, 74, 051104, 2008.

\bibitem{SWPRL}
S.~Boettcher, B.~Gon{\c c}alves, and H.~Guclu.
\newblock {\em J. Phys. A: Math. Theor.}, 41:252001, 2008.

\bibitem{citeulike:99}
D.~J. Watts and S.~H. Strogatz.
\newblock {\em Nature}, 393:440, 1998.

\bibitem{Boccaletti06}
S.~Boccaletti, V.~Latora, Y.~Moreno, M.~Chavez, and D.-U. Hwang.
\newblock {\em Phys. Rep.}, 424:175, 2006.

\bibitem{SWN}
S.~Boettcher and B.~Goncalves.
\newblock {\em Europhys. Lett.}, 84:30002, 2008.

\end{thebibliography}

\end{document}